# Photoacoustic methane detection assisted by a gas-filled anti-resonant hollow-core fiber laser

**Cuiling Zhang,**[a] **Jose Enrique Antonio-Lopez,**[b] **Rodrigo Amezcua-Correa,**[b] **Yazhou Wang,**[a] **and Christos Markos,**[a,b,*]

[a]DTU Electro, Technical University of Denmark, 2800 Kgs. Lyngby, Denmark
[b]CREOL, The College of Optics and Photonics, University of Central Florida, 32816 Orlando, Florida, USA
[c]NORBLIS ApS, Virumgade 35D, 2830 Virum, Denmark

**Abstract:** Photoacoustic spectroscopys (PAS)-based methane ($CH_4$) detectors have garnered significant attention with various developed systems using near-infrared (NIR) laser sources, which requires high-energy and narrow-linewidth laser sources to achieve high-sensitivity and low-concentration gas detection. The anti-resonant hollow-core fiber (ARHCF) lasers in the NIR and mid-infrared (MIR) spectral domain show a great potential for spectroscopy and high-resolution gas detection. In this work, we demonstrate the generation of a frequency-comb-like Raman laser with high pulse energy spanning from ultraviolet (UV) (328 nm) to NIR (2065 nm wavelength) based on a hydrogen ($H_2$)-filled 7-ring ARHCF. The gas-filled ARHCF fiber is pumped with a custom-laser at 1044 nm with ~100 µJ pulse energy and a few nanoseconds duration. Through stimulated Raman scattering process, we employ the sixth-order Stokes as case example located at ~1650 nm to demonstrate how the developed high-energy and narrow-linewidth laser source can effectively be used to detect $CH_4$ in the NIR-II region using the photoacoustic modality. We report the efficient detection of $CH_4$ with sensitivity as low as ~550 ppb with an integration time of ~40 s. In conclusion, the main goal of this work is to demonstrate and emphasize the potential of the gas-filled ARHCF laser technology for compact next-generation spectroscopy across different spectral regions.

## 1 Introduction

Trace gas detection is vital in various fields, such as industrial process monitoring, atmospheric and environmental research, and exhaled breath analysis [1-3]. Several methods have been proposed for low concentration gas detection, such as second harmonic wavelength modulation spectroscopy (2f-WMS) [4], photoacoustic spectroscopy (PAS) [5-7], and cavity enhanced absorption spectroscopy (CEAS) [8] to name a few. Among various detection principles, PAS offers notable advantages such as high sensitivity, low detection limit, and compact configuration. The PAS-based method is based on "listening to" the acoustic signals excited from the optical absorption of the targeted gas molecules. Methane ($CH_4$), as a major gas in atmosphere and greenhouse emissions, is crucial for real-time monitoring in natural and industrial settings and requires an optical detection with high selectivity and accuracy in complex atmospheric gas mixtures. Recently, this gas has attracted significant attention from the community attempting to develop PAS-based $CH_4$ detectors using different lasers in the near-infrared (NIR) wavelength range (See Table 1 in Appendix A).

NIR wavelengths are widely preferred compared to the MIR because it is a mature spectral region with available components. Commonly used lasers for PAS detection of $CH_4$ include the distributed feedback diode laser (DFB-DL) [9-12] and optical parametric oscillator (OPO) [12]. However, these laser sources come with certain limitations related to portability, beam quality, and robustness. Since the amplitude of the acoustic signal is proportional to pulse energy, the higher the excitation pulse energy, the higher sensitivity and lower detection limit [8] can be obtained. The narrow absorption line of $CH_4$ also requires a narrow-linewidth laser source to avoid interference from other gas species. Besides, although with narrow linewidth, the laser diodes operate at modulation mode and therefore the PA detection is sensitive to environmental temperature variations, compromising the reliability and selectivity [13]. Currently, this kind of pulsed laser with high energy, narrow linewidth and cross-sensitive free targeting the desired absorption line with a compact structure is still a challenging task.

Recent developments in anti-resonant hollow-core fiber (ARHCF) lasers have introduced a novel dimension to the landscape of laser sources, particularly in diverse spectral application domains such as gas spectroscopy and biomedical imaging. [13]. The intrinsic stimulated Raman scattering (SRS) characteristics of gas medium combining with the low-loss and high-damage-threshold ARHCF technology constitute a promising way to flexibly convert wavelength over a broad range to form the high-energy and narrow-linewidth Raman pulses [14-19]. In comparison to the aforementioned lasers, gas-filled ARHCF lasers offer a wide spectral range, spanning from the ultraviolet (UV) to the MIR, high pulse energy, narrow linewidth and compact size [20-23]. Therefore, their unique performance makes gas-filled ARHCF laser technology a competitive light source for spectroscopic applications [13, 24]. However, while this laser technology has existed for several years, there are still to be employed in real-life applications such as PAS of $CH_4$.

This work investigates the use of a frequency-comb-like multiline laser based on a hydrogen ($H_2$)-filled ARHCF and shows the feasibility to utilize the generated Stokes Raman line for photoacoustic modality to detect $CH_4$. We have shown a comb-like laser source that spans from UV to NIR generated when the $H_2$-filled ARHCF is pumped at 1044 nm and 1060 nm with a pulse duration of ~3.7 ns and average energy of ~100 μJ. Then the sixth-order rotational Raman Stokes (RRS) at ~1650 nm pumped at 1044 nm is adopted as excitation for photoacoustic detection of $CH_4$. The stability and repeatability are also analyzed for evaluation of the performance of gas detection.

## 2 Experiment Setup

The system consists of two parts which are: the excitation source based on a 7-tube $H_2$-filled ARHCF laser and the photoacoustic gas detection system, as shown in Fig. 1. The pump laser consists of a modulated diode laser seed followed by an Yb-doped fiber amplification module. The pump laser delivers a linear polarized pulse train at 1 kHz repetition rate with ~3.7 ns pulse duration and ~98 µJ pulse energy [13, 25]. This custom-made laser provides two switchable wavelengths (1044 nm and 1060 nm) that can be precisely tuned by changing the temperature of the diode in the range of 20°C to 35°C. This property is critical for high selectivity $CH_4$ sensing because the absorption band of gases are mainly composed of a series of discrete absorption lines. The output beam of the pump laser is collimated and coupled into a 5-m long ARHCF with a coupling efficiency of ~80%. A quarter-waveplate is placed before the ARHCF to optimize the polarization orientation and thus enhance the rotational SRS effect. Two high pressure gas cells developed in house are used to seal the ARHCF and to provide access for light coupling through $CaF_2$ windows. The ARHCF consists of 7 rings with a diameter of 16.1 μm and wall thickness of ~323 nm, forming a negative-curvature core shape with an inner jacket tube diameter of 32.8 μm, which shows a simulated transmission loss of <0.5 dB/m from 1 μm to 1.9 μm [13, 25] and as low as 0.13 dB/m at ~1650 nm. The simulation model can be found in [26, 27].

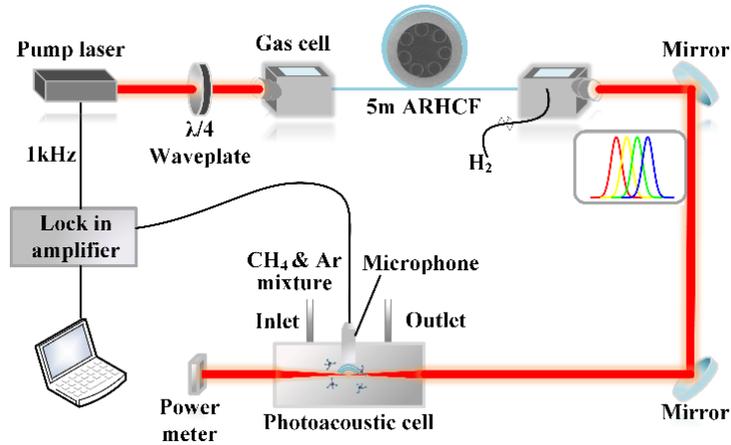

**Fig. 1 Experiment setup of the developed laser and photoacoustic system.** ARHCF: anti-resonant hollow-core fiber.

The generated output Raman laser beam is collimated and then is coupled into an in-house-built PA cell, where light interacts with the gas sample to be detected. The structure and fabrication of the PA cell have been described in our previous work [13, 23]. A microphone (4955, Brüel & Kjær) is used to collect the PA signal excited inside the PA cell. Diluted $CH_4$ with 1000 ppm concentration is connected to two mass flow controllers with 10 sccm flow rate, and pure Argon (Ar) used for diluting $CH_4$ is connected to another MFC with 2000 sccm flow rate. This dilution system allows for regulating $CH_4$ with different concentrations. The PA signals collected by microphone are processed by a lock-in amplifier (MFLI 500 kHz, Zurich Instruments). The PA cell is sealed with $CaF_2$ window, and the output power after PA cell is recorded by a power-meter to monitor the stability of the laser.

## 3  Results and discussion

*3.1 Characterization of the frequency-comb-like Raman laser*

Employing the filled $H_2$ with a RRS coefficient of 587 $cm^{-1}$ [28], the frequency-comb-like Stokes lines ranging from 328 nm to 2065 nm when pumped at the 1044 nm wavelength are generated, as presented in Fig. 2(a). Similar frequency-comb-like Stokes lines can be generated by pumping laser at 1060 nm wavelength, but with different conversion wavelength ranging from 421 nm to 2110 nm instead. Here since the spectrum spans from ~0.3 μm to 2 μm, three different optical spectrum analyzers (OSA) were utilized to record the spectra at different wavelength ranges (Spectro 320 Instrument Systems, Yokogawa AQ6375, and AQ6317B). The spectrum contains strong RRS lines in NIR wavelength range and weak anti-Stokes lines in the UV and visible wavelength range. Four octaves Raman combs can be observed, including the Raman line at ~1650 nm for $CH_4$ detection. We can see strong 8-orders of rotational Stokes lines from 1100 nm to 2065 nm. The spectrum that spans less than 1 μm wavelength is shown in the logarithmic form in the inset for clear illustration, where the pulse energy is lower compared with the NIR region. Two vibrational Raman lines at ~450 nm and ~528 nm are generated with several relatively weak rotational sidebands by properly changing the orientation of the polarization direction. Although the energy in UV and visible is weaker than the NIR range, we can see two vibrational anti-stokes Raman lines with rotational sidebands from 300 nm to 1000 nm when the gas pressure is increasing to ~15 bar (Fig. 7 in Appendix B). The generated weak energy at short wavelengths less than 1 μm is due to the phase mismatching for Raman anti-Stokes generation.

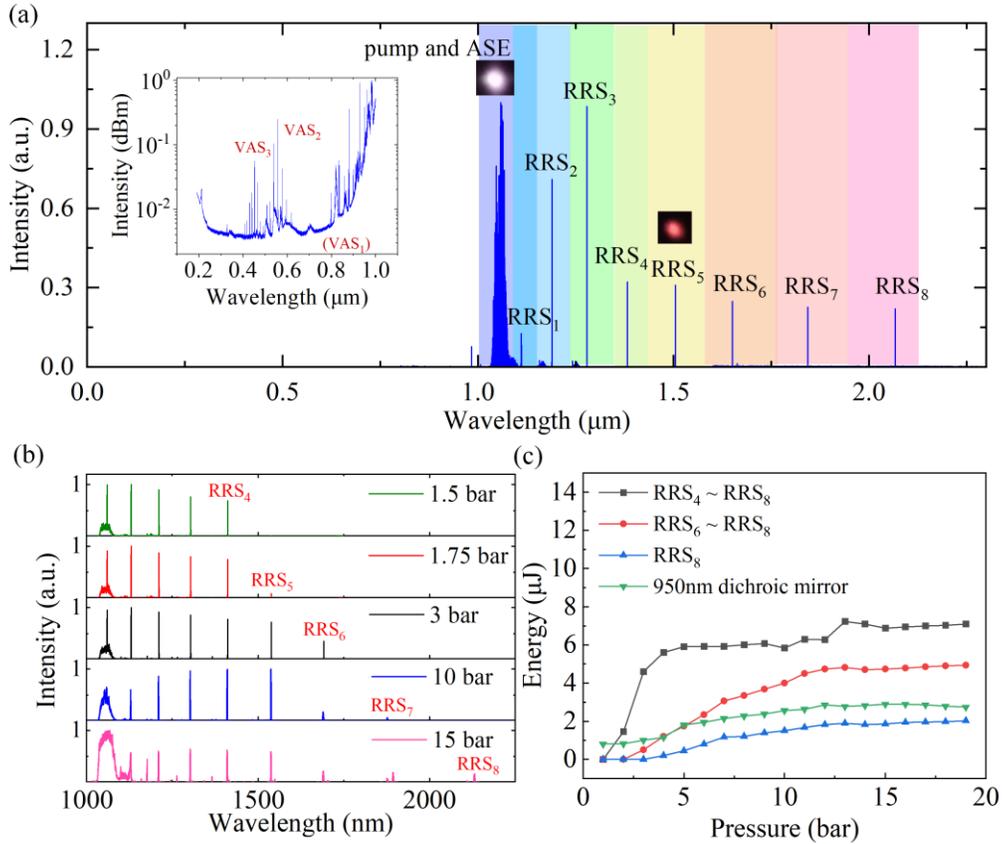

**Fig. 2.** (a) Spectrum from UV to NIR and spots of Raman lines. (b) Spectrum evolution of NIR wavelength at 1.5 bar, 1.75 bar, 3 bar, 10 bar, and 15 bar pressure. (c) Pulse energy of Raman laser against H$_2$ pressure.

Since the sensitivity and detection limit strongly depend on the pulse energy, the enhancement of energy is vital. The spectrum evolution of NIR wavelength range as well as the energy evolution were recorded in Fig. 2(b) and (c). Here the spectrum is from 1060 nm pump laser, but the spectrum evolution process is similar with 1044 nm pump. Initially, only 1st to 4th RRS lines can be observed at 1.5 bar pressure as shown in Fig. 2(b). As the $H_2$ pressure increases, the Raman gain coefficient increases, and another 5th-order Stokes line is formed from the high-energy RRS line at 1.75 bar. Successively, the 6th-order to 8th-order Stokes line can be formed from the generated Stokes lines at 3, 10, and 13 bar pressure, respectively. The generation of the next Stokes line follows an energy decrease of the former Stokes line. When the pressure increases to 20 bar, the energy of the 8th-order Stokes line gradually saturates because the 9th-order Stokes line cannot be generated due to the relatively high fiber loss at ~2.4 μm wavelength. Different filters were used to measure the power of different Raman lines in the NIR region. Figure 2(c) shows the energy evolution as a function of the gas pressure indicating that the pulse energy of the 1650 nm Raman line reaches the highest value of 2~4 μJ at ~20 bar, which is sufficient for the excitation of acoustic signals from the methane. The Gaussian-like beam profile of $RRS_6$ at 1650 nm, measured by a beam profiler (BP109-IR2, Thorlabs)) in the inset of Fig. 2(a) indicates that the laser operates in the fundamental mode of the fiber.

*3.2 Photoacoustic $CH_4$ concentration detection*

Here we employed the ~1650 nm Raman Stokes line for detecting $CH_4$. The wavelength of the Raman line can be precisely thermally tuned by tuning the laser diode seed at ~1044 nm in the temperature range of 20℃ to 35℃. This is critical for $CH_4$ detection because it is known that the absorption band of $CH_4$ gas is composed of a series of discrete narrow lines, therefore requiring the wavelength of the Raman line to be accurately overlapped with one of these absorption lines. Figure 3(a) shows the precise tunability of the wavelength of the 6th-order RRS line, recorded by OSA (Yokogawa AQ6375) with a step size of 0.024 nm. Although the linewidth cannot be accurately resolved due to the limited resolution of OSA, the Raman laser has a narrow linewidth

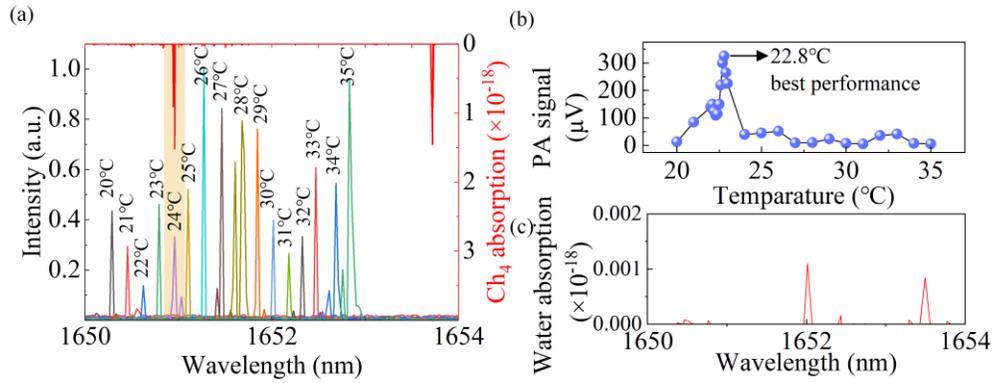

**Fig. 3** (a) Absorption spectrum of $CH_4$ (from HITRAN [29]) and laser spectra at ~1650nm at different laser temperatures. (b) PA signal against laser temperature. (c) Absorption spectrum of water (from HITRAN [29]).

of less than 0.1 nm, which is an advantage to achieve a high sensitivity. From the spectrum in Fig. 3(a), we find the wavelength at ~24°C is the best for overlapping with the one of the absorption lines of $CH_4$. To obtain the precise temperature for the optimal PA detection of $CH_4$, the amplitude of PA signal was recorded and compared by precisely tuning the wavelength of the 1650 nm Raman line. In this experiment, 1000 ppm $CH_4$ keep flowing through the PA cell at a flow rate of 300 sccm. Figure 3(b) shows the measured result, where it can be seen that the laser diode operating at 22.8°C excites the highest PA signal amplitude, therefore, in this experiment we select 22.8°C as the optimal temperature. Furthermore, the selected wavelength can also avoid the interference from the water (vapor) introduced by the high-peak power of the pulsed laser, due to the low absorption of water (Fig. 3(c)).

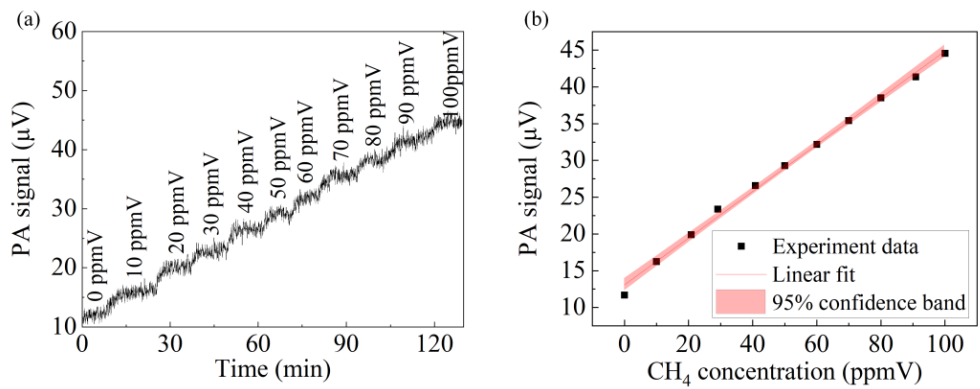

**Fig. 4 Gas detection results and system characterization.** (a) Photoacoustic signal against $CH_4$ concentrations from 0 ppm to 100 ppm. (b) Linear model between photoacoustic signal and concentration.

Figure 4 shows the $CH_4$ detection results including the linear fitting between the different concentrations and PA signal amplitude. The PA signals at different concentrations ranging from 0 to 100 ppm with an interval of 10 ppm were recorded and shown in Fig. 4(a). This is a continuous measurement process within 2.2 h where the transition time between two adjacent concentrations is around 10 minutes and is mainly determined by the gas diffusion rate within the dilution system and the PA cell. As shown in Fig. 4(b), the amplitude of the PA signals at different concentration levels shows a good linear relation with the concentration of $CH_4$.

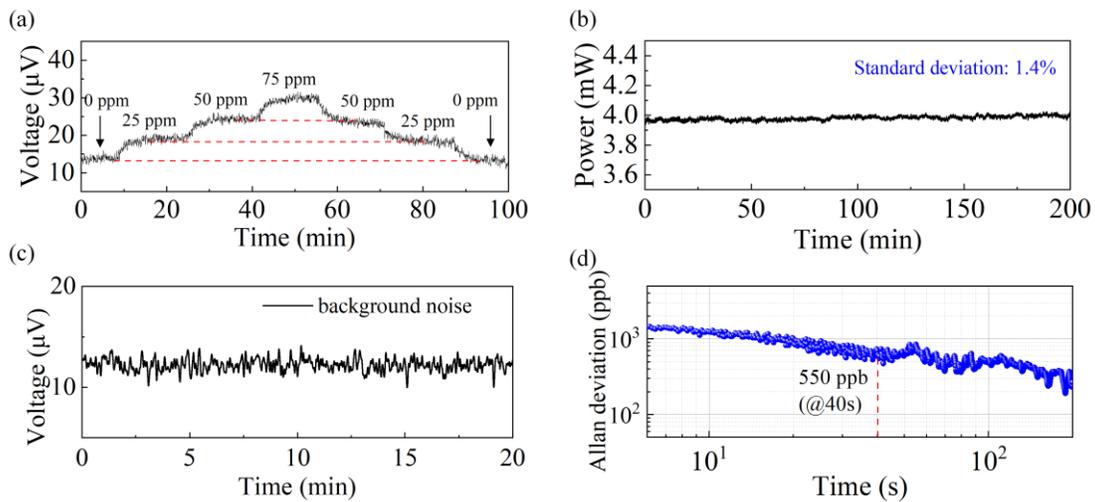

**Fig. 5 Gas detection system characterization.** (a) Repeatability of gas detection within 100 minutes. (b) Power monitoring of beam after photoacoustic gas cell with pure Ar within 200 minutes. (c) Background noise signal recorded with pure Ar within 20 minutes. (d) Allan deviation of background signal in (c).

To evaluate the overall performance of the system, we characterized its repeatability and stability over several hours. In Fig. 5(a), the PA signals of increasing and decreasing the concentration were monitored continuously, where concentration of 0 ppm, 25 ppm, 50 ppm, and 75 ppm were monitored, respectively. The consistency of the PA signal amplitude at the same concentration levels when increasing and decreasing the gas concentrations shows great repeatability within 100 minutes. Figure 5(b) presents the pulse power of the beam after the photoacoustic gas cell when the concentration of $CH_4$ is 0 ppm by only filling pure Argon to the gas cell. There is no obvious shift of the power within around 200 minutes with a standard deviation of ~1.4%, showing the stability of laser. As shown in Fig. 5(d), the Allan deviation of the background noise was calculated to evaluate the noise performance and detection limit based on the background noise signals in Fig. 5(c). The sampling interval of the background noise signals is set to be 0.3 s, and 4000 points are recorded within 20 minutes to calculate the Allan deviation. This decrease is a sign that the background noise mainly obeys a white noise distribution, and the drift is negligible [13, 23, 30]. The detection limit is calculated to be ~7 ppm using data of the standard deviation of the background signal and the linear model in Fig. 4(b) [30]. The sensitivity of the system was further evaluated from Allan deviation. The linear drop in the logarithmic scale with the increase of integration time indicates that lower detection limit can be achieved by increasing the lock-in integration time. The minimum sensitivity of our system was determined to be ~550 ppb with integration time of 40 s.

*3.3 Discussion*

The gas-filled ARHCF laser shows the potential in $CH_4$ detection by providing a tunable and narrow linewidth Raman line, with a minimum detection limit to be ~550 ppb with an integration time of 40 s. The current work suggests the development of a compact laser source with high beam quality to achieve improved coupling between light and acoustic and thus high sensitivity. The all-fiber structure is less sensitive to environmental conditions e.g., temperature and vibration. This work showcases an example for generating a Raman line for $CH_4$ detection. The detection limit can be further improved by enhancing the energy of the Raman line by changing the length of fiber and tuning the gas pressure. The gas-filled ARHCF laser can also be reconfigured to operate at

~3.4 μm wavelength targeting a higher absorption of CH$_4$ [31]. Also, the frequency-comb-like laser source can also be extended to the dual-gas detection of CH$_4$ and CO$_2$. As shown in Fig. 6(a), the 6$^{th}$-order RRS line at ~1695 nm generated by using the 1060 nm pump laser overlaps with one of the absorption spectrum lines of CH$_4$ while at the same time CO$_2$ has an absorption peak at around 2065 nm, which overlaps with the 8$^{th}$-order RRS line generated by using the 1044 nm pump laser. An initial investigation on CO$_2$ monitoring is investigated using the 2065 nm Raman line. The PA amplitude at different seed laser temperatures were recorded when filling 1000 ppm CO$_2$ in the gas cell (Fig. 6(b)), demonstrating the feasibility for CO$_2$ detection. In the future, it is also expected to achieve a real-time dual-gas detection of CH$_4$ and CO$_2$ by simultaneously switching the different Raman lines [13].

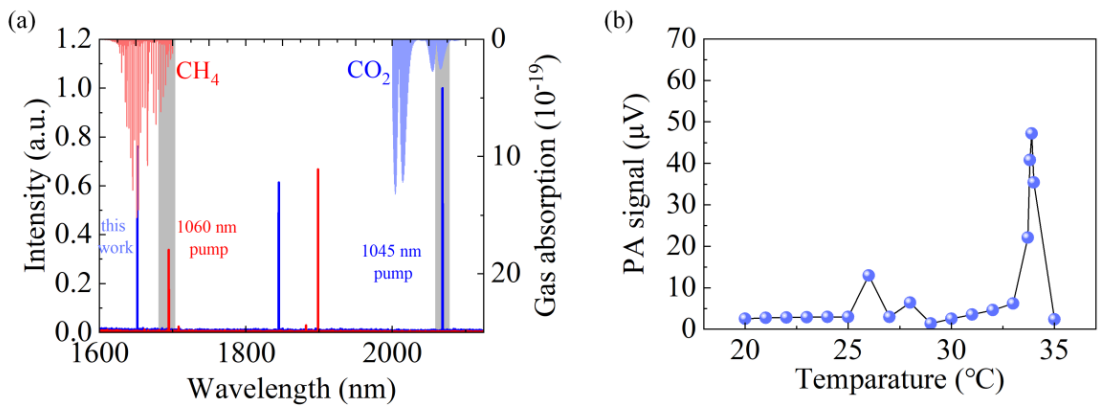

**Fig. 6** (a) Raman spectrum for dual-gas (CH$_4$/CO$_2$) detection generated from two pump lasers. Absorption spectrum of CH$_4$ and CO$_2$ are from HITRAN. (b) PA signal against the seed laser temperature for CO$_2$ detection.

## 4 Conclusion

In summary, the present work demonstrated a CH$_4$ concentration detection system enabled by the 6$^{th}$-order rotational Stokes line at ~1650 nm from a novel frequency-comb-like gas Raman laser source. We presented the characteristics of the frequency-comb-like laser spanning from UV down to 328 nm to extended NIR 2065 nm by a H$_2$-filled ARHCF and optimized the laser energy and wavelength for the CH$_4$ detection. Our research presents the employment of a high-pulse laser for CH$_4$ detection with a minimum limit of ~550 ppb and an integration time of 40 s. Improving the detection limit can be achieved if the pulse energy of the laser could be improved. This work also demonstrates the feasibility of generating high energy pulses at unique wavelengths by a series of frequency-comb-like Raman lines. Such a system can find use in a wide arc of applications in gas spectroscopy and label-free photoacoustic imaging [31].

*Appendix A*

Table 1 summarizes the State-of-the-art CH$_4$ detection results using PAS-based method using different laser sources in the NIR wavelength range. OPO: Optical parametrical oscillator; DFB: distributed feedback.

Table 1
State-of-the-art CH$_4$ detection results using PAS-based method.

| Ref | Laser source | Wavelength (nm) | Power (mW) | Detection limit (ppmV@integration time) |
|---|---|---|---|---|
| 1 [12] | OPO | 1650 | 60 | 0.020-0.030 |
| 2 [32] | InGaAs diode laser | 1630 | 3 | 40 |
| 3 [33] | Tunable DFB laser | 1651 | 10 | 0.5 (@1s) |
| 4 [34] | Diode laser | 1635.3 | 0.3 | 6 |
| 5 [10] | Laser diodes | 1650 | 1.5 | 1 |
| 6 [35] | Diode laser | 1650 | 0.7 | 60 |
| 7 [36] | DFB laser | 1650.96 | 20 | 0.009 (@500s) |
| 8 [37] | DFB laser | 1645.41 | 18 | 4 (@10s) |
| 9 [38] | DFB laser | 1650.9 | 12 | 8.4 (@1s) |
| 10 [39] | DFB laser | 1653 | / | 6.89 (@49s) |
| 11 [40] | Raman fiber amplifier | 1653.7 | 15.5 | 0.034 (@10s) |
| This work | Raman fiber laser | 1650 | 2~4 | 7 (@600ms) 0.550 (@40s) |

*Appendix B*

Figure 7 shows the spectrum evolution of wavelength range less than 1 μm from 12 bar to 20 bar pressure (with 1044 nm pump). At a lower pressure, two strong vibrational Stokes lines at 452 nm and 558 nm and weak rotational Stokes lines are generated. As the increase of pressure, more Stokes lines are formed at 15 bar pressure, and the shortest wavelength can reach 328 nm.

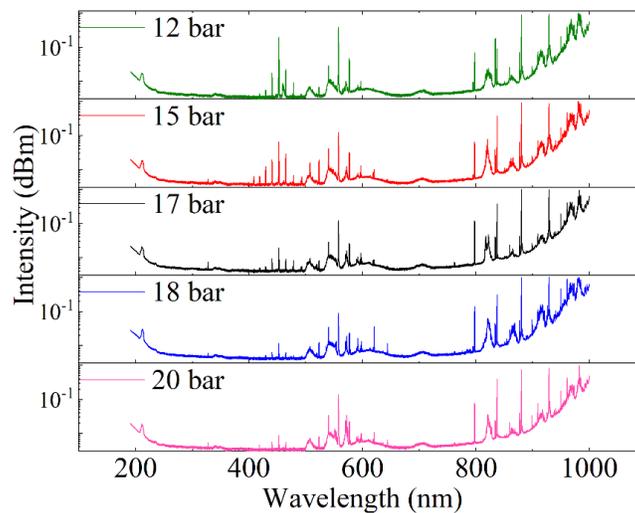

**Fig. 7** Spectrum evolution of wavelength range less than 1 μm at different gas pressure.

*Disclosures*

The authors declare that there are no conflicts of interest.

*Code, Data, and Materials*

The data of this work is available from the corresponding author upon reasonable request.

*Acknowledgments*

This work is supported by the VILLUM Fonden (Grant No. 36063, Grant No. 40964), and US ARO (Grant No. W911NF-19-1-0426).